# Advanced machine learning informatics modeling using clinical and radiological imaging metrics for characterizing breast tumor characteristics with the OncotypeDX gene array.

Michael A. Jacobs, Ph.D[1,2], Christopher Umbricht, M.D., Ph.D[1], Vishwa Parekh, MSE, Ph.D.[1,3],

Riham El Khouli M.D., Ph.D.[1], Leslie Cope, Ph.D.[4], Katarzyna J. Macura, M.D.,Ph.D. [1,2],

Susan Harvey M.D.[1], Antonio C. Wolff, M.D.[2]

[1]The Russell H. Morgan Department of Radiology and Radiological Science,
[2] Sidney Kimmel Comprehensive Cancer Center, and [3] Department of Computer Science,
[4]Department of Oncology.
The Johns Hopkins University School of Medicine, Baltimore, MD 21205

Address correspondence to:
Michael A. Jacobs
The Russell H. Morgan Department of Radiology and Radiological Science and Oncology,
Division of Cancer Imaging
The Johns Hopkins University School of Medicine,
Traylor Blg, Rm 309
712 Rutland Ave, Baltimore, MD 21205
Tel:410-955-7483
Fax:410-614-1948
email: mikej@mri.jhu.edu

**Short Title:** *Integrated Radiological Informatics System with comparison to Oncotype DX gene array*

*Initial results presented at the 2015 RSNA meeting, Chicago IL,*

This work was supported by the National Institutes of Health grant numbers: 5P30CA006973 (IRAT), U01CA140204, 1R01CA190299, Komen Scholar (SAC110053 (ACW), and Equipment donation of a K40 GPU card from the Nvidia Corporation.

*PREPRINT*

*Integrated Radiological Informatics System with comparison to Oncotype DX gene array*

**Abstract**

**Purpose**: Emerging data on breast cancer suggest that different breast cancer phenotypes may respond differently to available adjuvant therapies. Optimal use of established and imaging methods, such as multiparametric magnetic resonance imaging (mpMRI) can simultaneously identify key functional parameters and provide unique imaging phenotypes of breast cancer. Therefore, we have developed and implemented a new machine-learning informatic system that integrates clinical variables, derived from imaging and clinical health records, to compare with the 21-gene array assay, OncotypeDX.

**Materials and methods**: We tested our informatics modeling in a subset of patients (n=81) who had ER+ disease and underwent OncotypeDX gene expression and breast mpMRI testing. The machine-learning informatic method is termed Integrated Radiomic Informatic System (IRIS) was applied to the mpMRI, clinical and pathologic descriptors, as well as a gene array analysis. The IRIS method using an advanced graph theoretic model that transforms the patient space into a visualization heatmap and quantitative metrics. Summary statistics (mean and standard deviations) for the quantitative imaging parameters were obtained. Sensitivity and specificity and Area Under the Curve were calculated for the classification of the patients. Statistical significance was set at $p \leq 0.05$.

**Results**: The OncotypeDX classification by IRIS model had sensitivity of 95% and specificity of 89% with AUC of 0.92 with 19(low), 50 (intermediate), 12(high) patients classified in each risk group. The breast lesion size was larger for the high-risk group ($7.6 \pm 5.8 cm^2$) and lower for both low risk ($5.8 \pm 9.0 cm^2$) and intermediate risk ($4.6 \pm 5.4 cm^2$) groups. There were significant differences in PK-DCE and ADC map values in each group. The lesion ADC map values for high- and intermediate-risk groups were significantly lower than the low-risk group ($1.09$ vs $1.38 \times 10^{-3} mm^2/s$).

**Conclusion**: These initial studies provide deeper understandings of imaging features and molecular gene array OncotypeDX score. This insight provides the foundation to relate these imaging features to the assessment of treatment response for improved personalized medicine.

*Integrated Radiological Informatics System with comparison to Oncotype DX gene array*

**Introduction**

Integrating clinical health information with imaging as well as other biomarkers could be beneficial for defining variable cancer phenotypes. This accurate integration of what now is seemingly disparate data may improve our understanding of the complex nature of cancer. Ultimately, this integration may provide predictive markers with clinical benefits in certain cancer phenotypes. For example, in breast cancer, there is active research on how to predict the potential of local recurrence after conservative treatment. One such method to predict recurrence of the cancer is the Oncotype DX assay. Oncotype DX is based on the mRNA expression by RT-PCR for estrogen receptor (ER) positive disease without the human growth factor receptor 2 (HER2-nu) overexpression and is tested on tissue obtained at biopsy or surgical samples [1,2]. Oncotype DX has been validated in prospective-retrospective studies as a prognostic tool in ER-positive patients treated with tamoxifen. It has been shown to be a predictive tool to identify patients most likely to benefit from the addition of adjuvant chemotherapy to endocrine therapy [1,3-5]

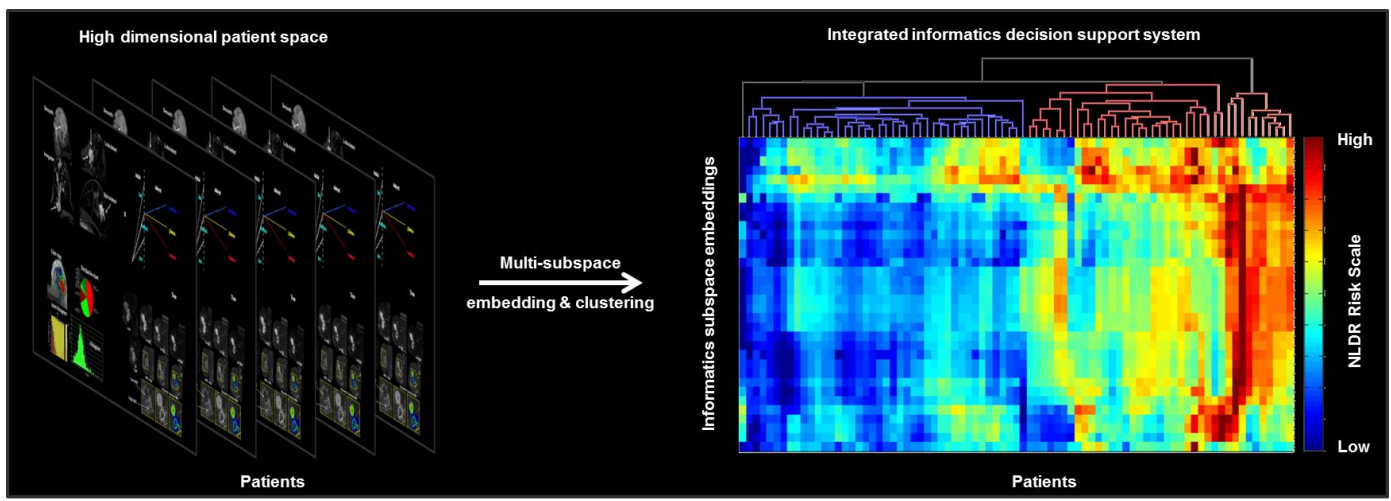

**Figure 1**. Illustration of the multi-subspace embedding and clustering method. The high dimensional patient space (left) consisting of different patients and their corresponding clinical and imaging parameter information is transformed into an integrated radiomics informatics system (IRIS) decision support system (right) using the multi-subspace embedding and clustering method. The IRIS results are represented using a heatmap where color scale (blue – red) indicates risk identified by each embedding while the hierarchical tree structure indicates the final patient classes (low, intermediate and high-risk)

Multiparametric (mp) radiological imaging can accurately detect and characterize breast lesions using advanced quantitative parameters [6,7]. For example, using dynamic contrast-enhanced (DCE)-MRI the vascularity and permeability of malignant lesions are characterized by rapid uptake of contrast agent followed by fast washout. Moreover, intra- and inter-cellular tumor environments of breast lesions are characterized by diffusion-weighted imaging (DWI), with the apparent diffusion coefficient (ADC) of water map. The ADC map provides a quantitative biophysical parameter that measures the cellularity of a lesion. The challenge is to accurately combine mpMR images with clinical and pathologic features to stratify patients and identify the potential for cancer recurrence, similar to the costly Oncotype DX. To answer this challenge, we have developed a new machine-learning informatic method termed Integrated Radiomic Informatic System (IRIS), which can be applied to multiparametric MRI, clinical and pathologic descriptors, as well as a gene array analysis[8,9]. The purpose of this study is to test the IRIS algorithm by combining data from imaging and clinical health records to stratify patients into three risk groups: low, medium, and high risk and then compare with the OncotypeDX 21-gene array assay outcomes.

**Methods**

**Clinical Subjects:**

All studies were performed in accordance with the institutional guidelines for clinical research under a protocol approved by our Institutional Review Board (IRB) and all HIPAA agreements were followed for this retrospective study. Patients were selected from the Johns Hopkins Integrated Breast Cancer Research Database developed by one on the authors and underwent MRI as part of the clinical health record review. Of the patients with biopsy proven breast cancer who presented to our facility for bilateral breast MRI, 123 patients were identified to have both the OncotypeDX and an advanced MRI exam which included DCE and DWI. Our inclusion criteria were: 1) breast imaging on 3T MRI scanner,

2) DCE and DWI MRI sequences, and 3) pathology proven diagnosis of ER+ breast cancer and 4) OncotypeDX having been performed on lesion tissue samples. There were 81 patients with 84 lesions that satisfied the inclusion criteria.

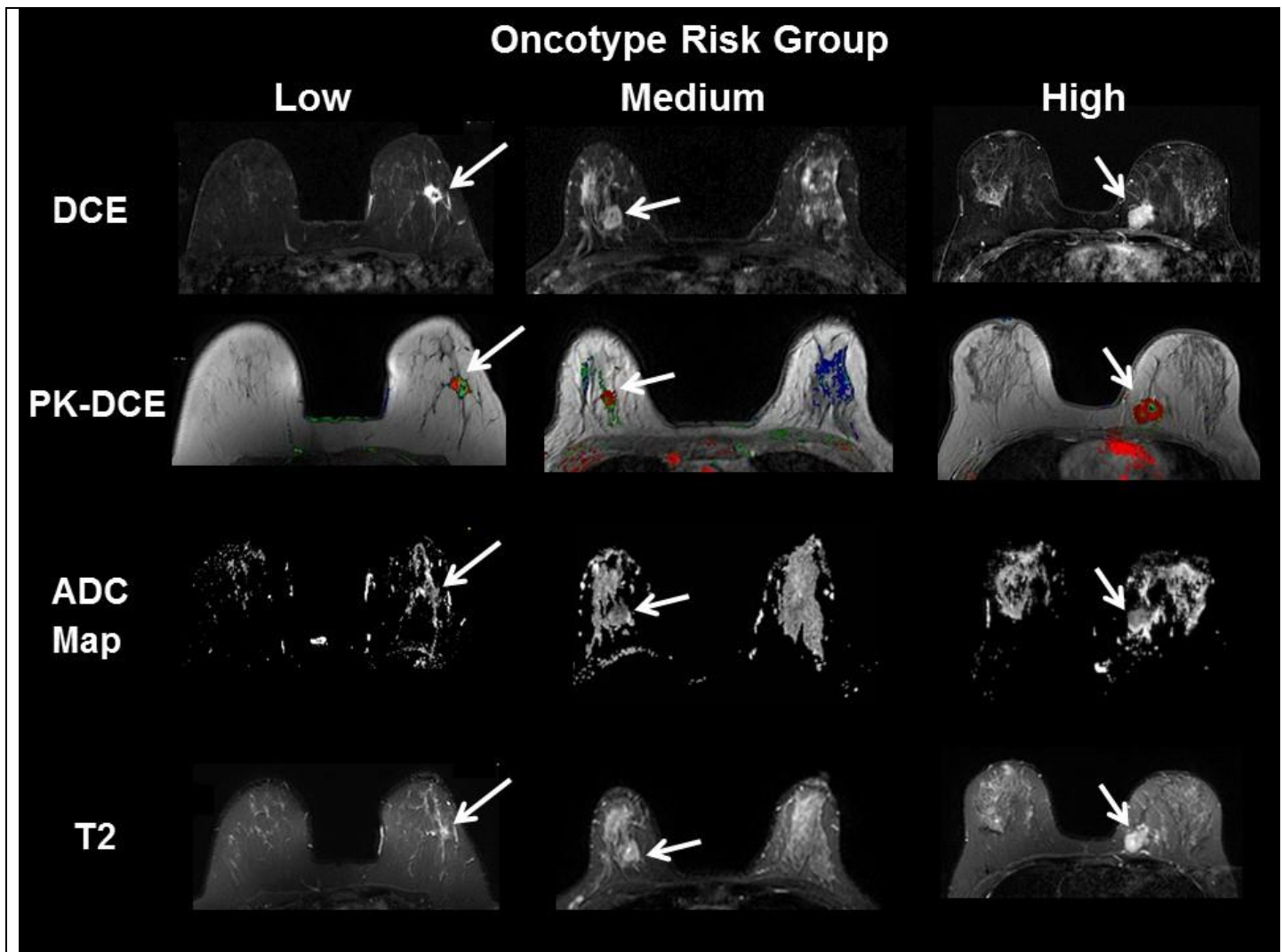

Figure 2. Demonstration of multiparametric breast MRI imaging of each risk group defined by the Oncotype DX. **Left Column**) typical imaging of the low risk patient. **Middle Column**) typical imaging of the medium risk patient, and **Right Column**) typical imaging of a high-risk group patient. Note, the PK-DCE all demonstrate malignant phenotype, however, by integrating all the data using IRIS, we were able to separate each Oncotype DX group.

**Histological Phenotyping:** All breast cancers were categorized by histological phenotyping based upon immunohistochemistry(IHC). Estrogen and progesterone receptors (ER and PR), HER2-Nu by FISH, and Ki-67 proliferation index (%). The Elston tumor grades for each lesion were distributed as Grade 1 (9%), Grade 2 (78%), and Grade 3 (13%), Histopathological data was obtained from the breast pathology database. All patient demographics matched the current clinical criteria for the Oncotype DX test (ER positive).

**Theory: Multi-subspace embedding and clustering:**

We have developed a machine learning model that integrates different types of clinical and imaging parameters, which allows for the construction of a clinical support decision model. The inherent high dimensionality of the clinical and imaging parameters and any complex correlations within the data presents significant challenges for integration and visualization of the data. These challenges were solved using nonlinear dimensionality reduction (NLDR) method [10]. Nonlinear dimensionality reduction algorithms transform and embeds a *D* dimensional space into a lower *d* dimensional manifold representation of *D's* intrinsic dimensionality, where *d<D*. The goal of the multi-subspace embedding and clustering method is to transform the patient space, represented as $X = \{x_1, x_2, ..., x_n\} \subset R^D$ where, $x_i$ represents the i[th] patient, n represents the number of patients and D represents the number of clinical and imaging parameters, into an integrated radiomics informatics decision support system (IRIS) visualized by a heatmap as shown in **Figure 1**. The steps of the IRIS system are outlined below[8,9].

In the first step, a *D* dimensional patient space is transformed into *N* d-dimensional subspaces where $d \in \{2,3\}$. The second step involves transformation of each d-dimensional subspace, $S_i = \{x_1, x_2, ..., x_n\} \in R^d \ \forall i \in \{1,2, ..., N\}$ into a one dimensional embedding, $Y_i = \{x_1, x_2, ..., x_n\} \in R^1 \ \forall i \in \{1,2, ..., N\}$ using a nonlinear dimensionality reduction algorithm [11]. In the third step, each one-dimensional embedding, *Y<sub>i</sub>* is evaluated against the ground truth (OncotypeDX scale) based on the correlation coefficient, *Ri* between *Yi* and the ground truth. In this paper, we employed a simple evaluation metric of correlation coefficient. The aim of this step is to identify the optimal set of one dimensional embeddings, *U*, such that, $R_i \geq 0.5$, as shown in the following equation:

$$U = \{Y_i \in Y | R_i \geq 0.5\} \ \forall \ i \in \{1,2, ..., N\}$$

In the fourth step, a hierarchical clustering algorithm is used to cluster the set, *U* as well as the patient space spanned by the one-dimensional embeddings in *U* to produce two clustering configurations. The first clustering configuration between different one-dimensional embeddings in *U* provides us information about the relationships between the clinical and imaging parameter embedding, which can be used to identify the each parameter's importance, as well as remove redundant or similar embeddings. The second clustering configuration between each patient provides us input information about the relationship between different patients. This allows IRIS to identify patients clustered into different risk groups as well as classify any unknown patient into a relevant risk group.

**Clinical and imaging parameter importance:**

Using IRIS, the importance of each parameter, *i* is calculated by the percentage of embeddings in *U* that included the clinical and imaging parameter, *i*. The contribution of each parameter to the IRIS model allows for assessment of which parameters to keep and which to discard.

**Complex network analysis of informatics parameters**

a. Using IRIS, the high dimensional relationship between different clinical and imaging parameters was explored by modeling and analyzing a complex informatics network. Before modeling the complex network, the raw values corresponding to each clinical and imaging parameter were transformed into risk prediction score normalized in the range 0-1, such that zero corresponds to low risk and one corresponds to high risk. The risk prediction score for every parameter was calculated using the following steps: First, a correlation coefficient, $r_i$ between the values, $y_i$ spanned by the parameter, *i* across all the patients and corresponding Oncotype DX scores are calculated.

b. Second, the range of clinical and imaging parameter values across all the patients are normalized from zero to one according to the following formula

$$z_i = \begin{cases} \dfrac{y_i - \min(y_i)}{\max(y_i) - \min(y_i)}, & if\ r_i \geq 0 \\ 1 - \dfrac{y_i - \min(y_i)}{\max(y_i) - \min(y_i)}, & if\ r_i < 0 \end{cases}$$

Here $z_i$ represents the resultant risk prediction score for each parameter, *i*.

**Construction of network model of informatics parameters**

The complex informatics network, *G* was constructed from the multidimensional data, $Z = \{z_1, z_2, \ldots, z_K\} \in R^n$, where, $z_i$ represents the risk prediction score of the clinical and imaging parameter, *i*; *K* is the number of clinical and imaging parameters; and *n* is the number of patients. Furthermore, the complex network *G* is represented as G = (V,E) with $V = \{v_1, v_2, \ldots, v_K\}$ being the set of *K* vertices representing the clinical and imaging parameters and *E* being the adjacency matrix indicating the interactions between the clinical and imaging parameters in the form of edge weights. The edge weight between any two vertices $v_i$ and $v_j$ was computed as follows

$$E_{ij} = 1 - corr(z_i, z_j)$$

We defined a neighborhood parameter, k to define the number of nearest neighbors each parameter can be connected to. The connectivity in the resulting complex network is dependent on the value of k chosen. If the value of k is chosen to be too large, the complex network may produce short circuit or spurious connections while a low value of k would produce a disconnected network [11]. The value of k selected as 7, which approximates one-third of the total number of parameters.

**Statistical and topological characteristics of the complex network:**

The complex network was analyzed using graph summary metrics and centrality metrics [12,13] The average path length and diameter are the basic statistical metrics computed for any complex network. A path is defined as the set of edges connecting any two nodes and the sum of weights of these edges represent the path length. Average path length, as the name suggests, is the average of the path lengths across all pairs of nodes or clinical and imaging parameters. Diameter, on the other hand, is the maximum value among all the path lengths. Graph centrality metrics identify the most important clinical and imaging parameters in the complex network, also called the hub nodes [14,15]. These hub nodes influence the network properties. Furthermore, the probability of any incoming node connecting to these hub nodes is significantly higher than connecting to other nodes[13]. These hub nodes may correspond to the key clinical and imaging parameters that are predictors of breast cancer recurrence risk. In total, the following metrics were extracted from the complex network: Degree distribution, average path length, diameter, clustering coefficient and different centrality measures such as degree centrality, harmonics centrality, and betweenness centrality. (see appendix).

**Patient Classification**

We implemented patient classification using the hybrid IsoSVM feature transformation and classification algorithm [8] based on the Isomap [11] and the Support Vector Machine (SVM) algorithms [16] from the complex interaction network. The imbalance in the number of patients in different risk groups was overcome by setting different misclassification penalties for different risk groups while training the SVM. The optimal values for the Isomap neighborhood parameter and the misclassification penalty were estimated using leave-one-out cross validation.

**Multiparametric Breast Imaging**

Patients were scanned on a 3T MRI system (3T Achieva, Philips Medical Systems, Best, The Netherlands) using a bilateral, dedicated four-channel, phased array breast coil (InVivo, Orlando, FL) with the patient in the prone position. Representative multiparametric breast imaging for the three risk groups as defined by OncotypeDX are illustrated in figure 2.

**Proton MRI Imaging**: $T_2$-weighted spin echo (TR/TE/IR=7142/70/220ms, Field of View (FOV)=350x350, matrix=220×195, slice thickness(ST)=5mm, SENSE=2 and Averages(Ave)=2) and fast spoiled gradient echo (FSPGR) $T_1$-weighted (TR/TE=5.4/2.3ms, FOV=350x350, matrix=548×550, ST=3mm, SENSE=2 and Ave=1) sequences were acquired.

**Pharmacokinetic Dynamic Contrast-Enhanced MRI**. The Pharmacokinetic (PK) DCE was obtained using non-fat-suppressed(FS), three-dimensional(3D), FSPGR $T_1$-weighted (TR/TE=3.4/1.7ms,

FOV=350x350, matrix=256×126, Flip angle=10, slice thickness=5mm, and Ave=1) sequences. Pre- and fourteen post- contrast images (temporal resolution=15s) after intravenous administration via a power injector of a GdDTPA contrast agent (0.2 mL/kg (0.1mmol/kg)).

**High Resolution Dynamic Contrast-Enhanced MRI**. T1-weighted 3D GRE with FS(TR/TE=5.8/2.9ms, FOV=350x350, matrix=720×720, Flip angle=13, ST=3mm, and Ave=1) were obtained pre and post the PK-DCE.

**Diffusion-Weighted Imaging.** Diffusion-weighted imaging is acquired before contrast imaging using an FS fast spin echo EPI parallel imaging sequence (TR/TE/IR=9548/70ms, FOV=350x350, matrix=220×195, SENSE=2, Ave=2, ST=3mm, b=0,200,600,800 s/mm$^2$) on three planes and in less than three minutes. Trace apparent diffusion coefficient (ADC) maps were constructed using a monoexponential model from all three axes and interpolated to $256^2$.

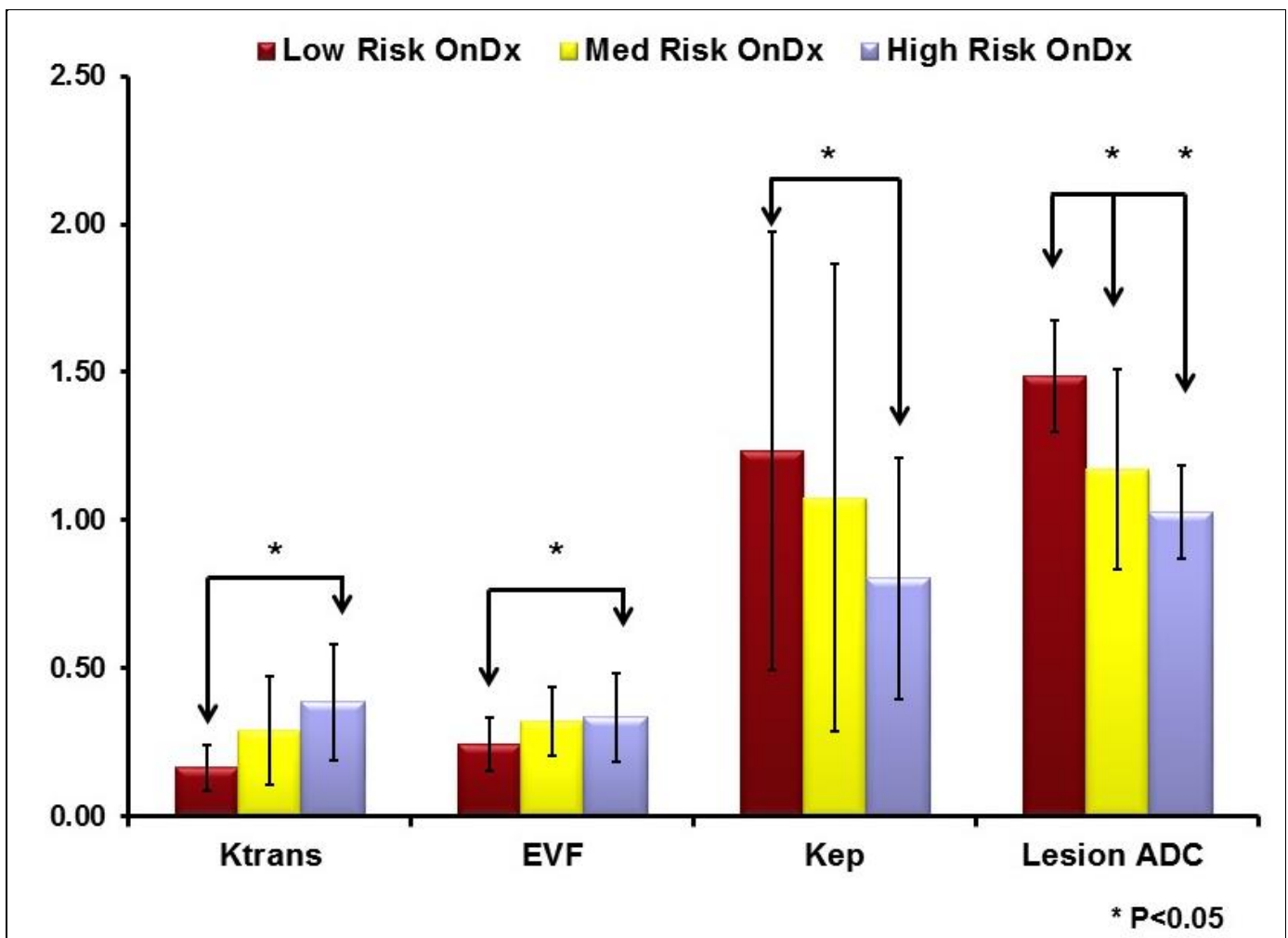

Figure 3. Bar graphs of quantitative multiparametric MRI parameters from the NLDR model. There are significant differences between each group of patients.

## MRI data analysis

**Clinical breast lesion classification methods:** Breast lesions were identified on the breast MRI by a radiologist who was blinded to pathological results defined by the BIRADS lexicon[17,18]. Breast density was defined as extremely dense tissue, heterogeneously dense tissue, scattered fibroglandular tissue or primarily fatty tissue. Background parenchymal enhancement (BPE) was defined as minimal, mild, moderate or marked. Lesions were classified as a focus, a mass or non-mass enhancement (NME). Morphologic assessment was defined for masses, as shape (round, oval, irregular) margins (1=circumscribed, 2=not circumscribed 2a= irregular 2b = spiculated) and enhancement patterns (1=homogenous, 2=heterogeneous or 3=rim). For NME, distribution (1=focal, 2=regional, 3=linear 4=diffuse, and 5=segmental) and enhancement pattern (homogenous, heterogeneous, clustered ring and clumped) were recorded. We defined lesion morphology into seven classes (1=focal NME, 2=regional NME, 3=linear or segmental NME, 4=circumscribed mass, 5=irregular mass, or 6=spiculated mass).

**Pharmacokinetic Contrast Enhancement Metrics:** Pharmacokinetic kinetic DCE MRI provides quantitative metrics of the volume transfer constant ($K^{trans}$ (min$^{-1}$)) which characterize uptake of the contrast agent, the leakage within the extracellular extravascular space ($v_e$ (%)), and the transfer rate constant ($k_{ep}$ (min$^{-1}$)). Post-processing of the DCE exam was performed by a combined Brix and Tofts model [19-21] using DynaCad (InVivo, FL) software from the identified breast lesions.

**ADC Mapping:** Regions of Interest(ROI) were drawn on normal appearing glandular tissue and breast lesions defined by DCE MRI. Means and standard deviations were calculated for both tissue types. Ratios of lesion ADC to glandular tissue ADC (L/GT) were calculated from the equation on lesion and glandular tissue [22].

$$\text{Normalized ADC value} = \frac{\text{ADC value of Lesion}}{\text{ADC value of glandular tissue}}$$

**Statistical Analysis**

We computed summary statistics (mean and standard deviations) for the quantitative imaging parameters from the mpMRI. An unpaired t-test was performed between each risk groups imaging parameters to determine statistical significance. Sensitivity and specificity and Area Under the Curve were calculated to determine the classification of the patients. Statistical significance was set at p ≤ 0.05.

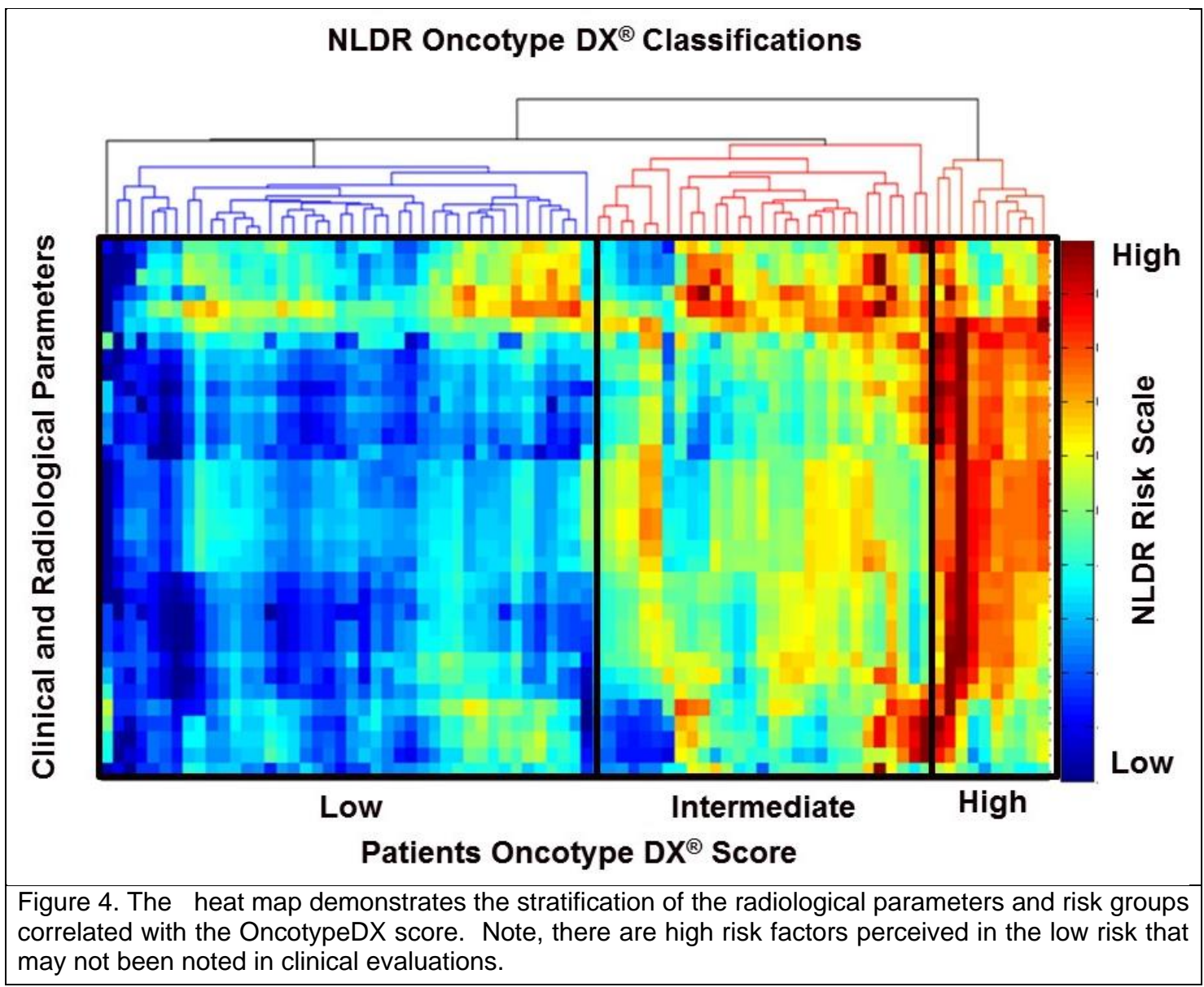

Figure 4. The heat map demonstrates the stratification of the radiological parameters and risk groups correlated with the OncotypeDX score. Note, there are high risk factors perceived in the low risk that may not been noted in clinical evaluations.

## Results

### Clinical Demographics:

81 patients with 84 lesions of 123 patients identified who had both multiparametric MRI imaging and Oncotype assay were selected. There were 19 (23%) patients with low risk (0-17), 50 (62%) patients with intermediate risk (18-31), and 12 (15%) patients with high risk (>31). Seventy-four patients were ER+/PR+ and seven had only ER+ expression. There was no age difference (51-56y/o) between the patients. These data are summarized in Table 1. The 42 patients not included in the study did not undergo a complete advanced mpMRI imaging session.

**Radiological Findings**

The high-risk group had the largest tumor size (7.6±5.8cm$^2$). Followed by the low risk group tumor size (5.8±9.0cm$^2$) and the intermediate risk group (4.6±5.4cm$^2$). For advanced MRI parameters, there were clear differences in each parameter and Oncotype risk groups. The PK-DCE parameters($K^{trans}$, EVF). The $K^{trans}$ values for the high- and intermediate-risk groups were higher (0.45 and 0.50 (1/min)) compared to the low-risk group (0.35(1/min)). Similar results were noted for the other PK-DCE parameters. the maximum contrast enhancement was largest for the high-risk group(523±145s), compared to the intermediate-risk(434±138s), and low-risk groups (489±139s). Similarly, the ADC map values from the high- and intermediate-risk patients in the lesion tissue were significantly lower (p<0.05) than those for the low-risk patients (1.09 vs 1.38x10-3mm$^2$/s). However, the ADC map values in glandular tissue remained constant across all groups (2.14-2.17x10-3mm$^2$/s). The bar graphs are shown in **Figure 3**

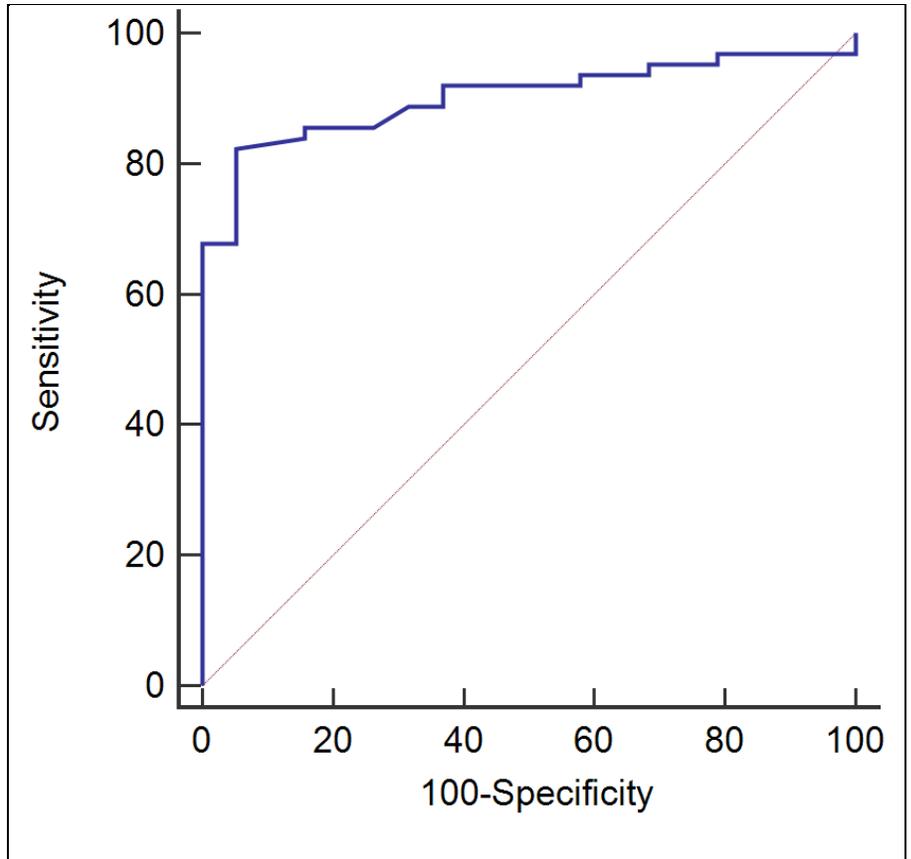

Figure 5. The Area under the Curve (AUC) graph of the top risk predictors in the NLDR model with the OncotypeDX score. The sensitivity was 95% and specificity was 82% with an AUC =0.92 (95% CI = 0.83-0.97).

**IRIS Model**

The IRIS heatmap demonstrating the risk profile of each patient is shown in **figure 4**. The IRIS system stratified the patients into three categories corresponding to low, intermediate and high-risk groups. The IRIS heatmap visualizes the individual contribution of each clinical and imaging parameter subspace using a color scale (low risk: blue to high risk: red). In addition, overall risk prediction of the patient is specified by the individual patient clusters. The top most important surrogate imaging and histological parameters determined from the clinical and imaging model are summarized in **Table 2**. These were the ADC map values, the PK-DCE parameters, and Ki-67. The NLDR embedding that produced the best area under curve (AUC) consisted of Ki-67, ADC lesion value and ADC ratio resulting in the AUC of 0.92 with sensitivity of 95% and specificity of 89%. The ROC curve is shown in **Figure 5**.

The topological graph theoretic metrics of integrated centrality and other different centrality measures for each informatics parameter are summarized in **Table 3**. The average path length between each parameter was 1.14, the correlation distance and diameter of the complex informatics network was 2.18. The average clustering coefficient was 0.53, much higher than the clustering coefficient of Erdos-Renyl random graph ($CC_{ER}$ = 0.0228). **Figure 6** illustrates the complex interaction network and the integrated centrality of each parameter between all the variables, which provides insight into the diverse relationship between them. The threshold on integrated centrality was chosen at 0.7 to determine the hub informatics parameters as there was a significant gap in the integrated centrality values between 0.7 and 0.54. The resulting hub informatics parameters included Ki-67, ADC lesion and ADC ratio, DCE kinetic curve type, and DCE pharmacokinetic parameters (Permeability, $K^{trans}$; extracellular extravascular fraction, EVF)

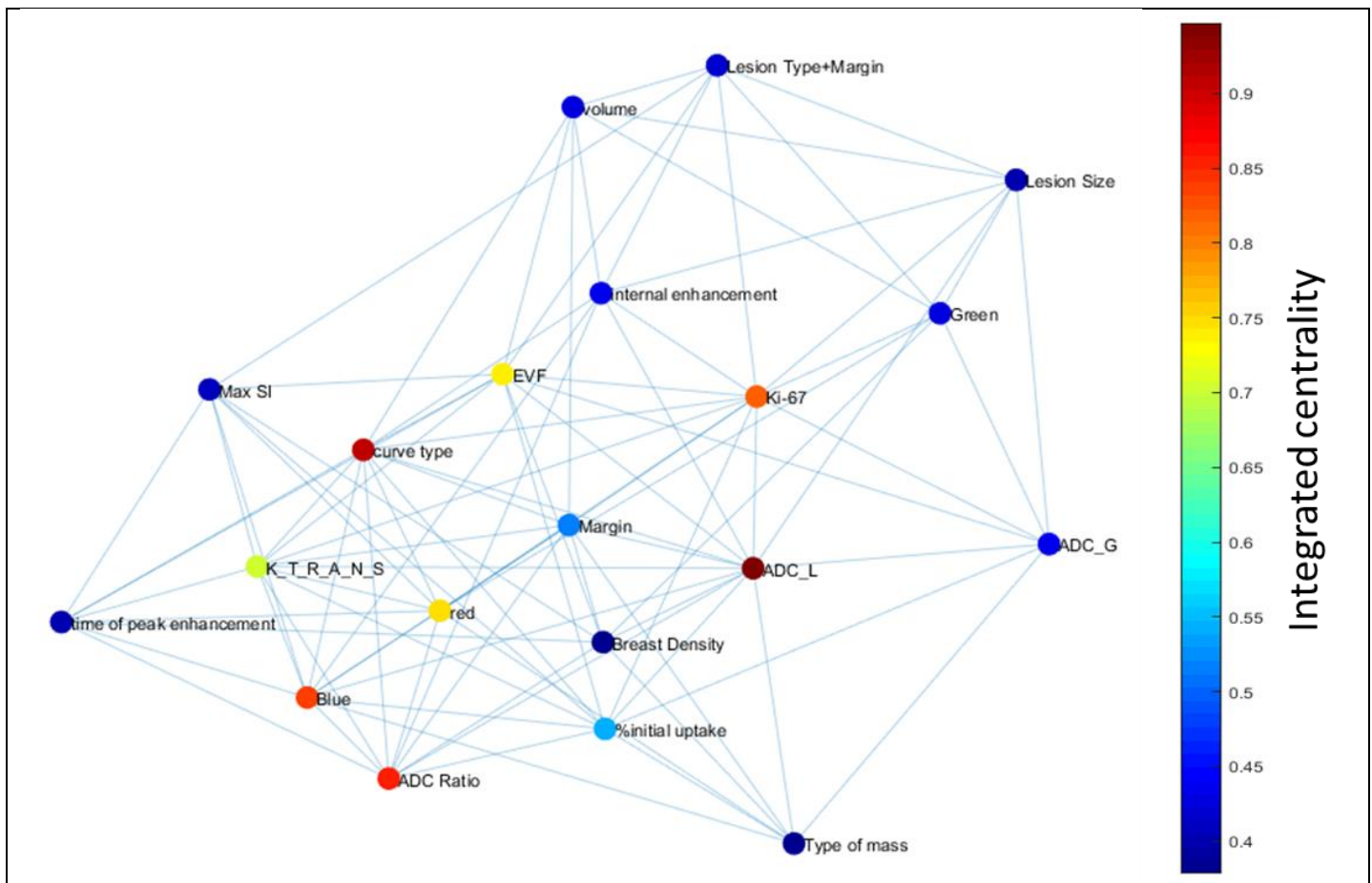

Figure 6. Visualization of integrated centrality on the complex network. The nodes were color coded from blue to red with blue representing minimum and red representing maximum integrated centrality values. The top IRIS parameters demonstrate high integrated centrality metrics demonstrating their importance.

**Discussion**

We have introduced and demonstrated an advanced NLDR integrated clinical and imaging model (IRIS) to analyze the relationships and interactions between mpMRI parameters, clinical heath records, and histological variables compared with the OncotypeDX assay. The IRIS model was able to consistently group patients into the three different categories based on data integration. Importantly, we defined imaging and clinical variables predictive of tumor recurrence compared to each category with the OncotypeDX. This separation of the data compared favorably to Oncotype DX and may lead to an accurate assessment for recurrence that could complement or provide data similar to those obtained with proliferation assays like Oncotype DX. For example, the most important parameters were ADC map values, PK-DCE metrics and Ki-67, all of which reflect the cellularity and vascularity of the tumor. This integrated model of imaging, clinical heath records, and IHC and histopathology data have the potential to describe features of cancer and may provide data for precision personalized care. This is the first study to employ an integrated graph theoretic model and machine learning using quantitative mpMRI and clinical variables in breast cancer compared with gene array data.

The categorization of the different risk groups from our model was strikingly consistent based on combined imaging and pathological IHC variables. Indeed, the ADC map values were lower in the high and intermediate risk group consistent with the increased Ki-67 from histological analysis. The PK-DCE parameters and lesion volume demonstrated similar characteristics. Moreover, this report demonstrates by using advanced unsupervised machine learning methods in breast cancer, that integration of several variables can accurately separate those cancers into different risk stratifications consistent with the OncotypeDX results. Finally, by using the complex interaction mapping, one can visualize the connections of each variable which may form the basis for even further predictive modeling.

The clinical and radiological parameters utilized in this study were derived from our clinical experience, since current treatment decision algorithms are based on standard clinicopathologic prognostic and predictive factors large datasets using clinical measures such as tumor size, node status, grade, ER, and HER2-nu [23-27]. Similarly, imaging features such as, breast density, lesion morphology, size, enhancement patterns as well as quantitative metrics (ADC map values and PK-DCE) are routinely used in practice and therefore were familiar to the radiologist and are readily available. Finally, the OncotypeDX is used a predictive tool to identify patients most likely to benefit from the addition of adjuvant chemotherapy to endocrine therapy and has been validated in prospective-retrospective studies [3-5]. Thus, the ability to combine these quantitative measures would be an important step in ensuring that "the right patient receives the right treatment".

We developed an integrated informatics decision support system (IRIS) based on multi-subspace embedding and clustering method for the purpose of diagnosis or prognosis. Furthermore, the complete multi-subspace embedding and clustering method is unsupervised and does not need any training data. The IRIS heatmap provides a visualization of relationship between different cancers along with a quantifiable embedding metrics. Using the IRIS heatmap, we would be able to identify a patient or a group of patients with the most similar informatics embedding metrics to a new patient with an unknown risk of recurrence. Understanding these complex relationships between different embeddings can provide an insight on how these embeddings are related at biological level predicting recurrence of breast cancer. Interestingly, the lesion size was not an accurate feature for categorizing cancers into the risk groups in this study. The lesion size was largest for the highest risk group, but was smallest for the intermediate risk group, suggesting that lesion size alone is not an accurate predictor. Differentiating and characterizing benign from invasive breast cancer is an important issue that was the focus of many different studies [7,28-32]. However, the prior studies typically only used a single or few MRI parameters for differentiating benign from invasive breast cancer. We employed machine learning and graph theory methods to differentiate these entities. Our IRIS model incorporated pathophysiological and imaging characteristics of different breast tissue types, enabling a more predictive model in the tumor environment. However, there some limitations. Our sample size was small, but based on known clinical and imaging variables and a proven gene array assay. This report provides initial data for further testing on a larger cohort. In addition, we plan on incorporating mammography and US into the model to provide additional data for improved classification and explore its use with Ductal Carcinoma in situ (DCIS).

In conclusion, these initial studies provide insight into the molecular underpinning of the surrogate imaging and clinical features and provide the foundation to relate these changes to histologic and molecular pathology parameters. The integration of these clinical and imaging parameters may help refine available prognostic and predictive markers, and improve clinical decision-making.

**ACKNOWLEDGMENTS**

This work was supported by the National Institutes of Health grant numbers: 5P30CA006973 (IRAT), U01CA140204, 1R01CA190299, Komen Scholar (SAC110053 (ACW), and Equipment donation of a K40 GPU card from the Nvidia Corporation.

**Appendix**

**Clustering coefficient:**

The clustering coefficient defines the connectedness of the neighborhood of an informatics parameter. Clustering coefficient ranges from zero to one with zero representing completely disconnected neighborhood and one representing completely connected neighborhood. Mathematically, the clustering coefficient, *i* is defined as

$$CC(i) = \frac{2e_i}{k_i(k_i - 1)}$$

Here $e_i$ is the number of connected edges neighborhood of *i*, $k_i$ are the number of nodes in the neighborhood of *i* making $\frac{k_i(k_i-1)}{2}$ the maximum possible number of edges in the neighborhood.

**Degree distribution:** The degree distribution is the most fundamental metric calculated for any complex network. The degree distribution, P(k) is determined using the following equation

$$P(k) = \frac{1}{N}\sum_{i=1}^{N} \mathbf{1}\{\deg(i) = k\} \ \forall k \in \{7, 8, \ldots, N-1\}$$

Here, deg*(i)* represents the degree of the parameter, *i* which is defined as the total number of parameters that are directly connected to it; N is the number of parameters.

The degree distribution enables us to identify whether the network is a scale-free network or not. For scale free networks, the degree distribution follows the power law $P(k) \propto k^{-\gamma}$, with the value of $\gamma$ ranging between 2 and 3. Scale free networks are characterized by the presence of few highly connected hub nodes that influence the network properties and may correspond to the key parameters that are predictors of breast cancer recurrence risk [13].

**Centrality measures:** The Centrality measures determine the importance of each informatic parameter in the complex network. The most widely used measures of centrality are betweenness, Harmonic and Degree centrality [14,15,33]. We use three centrality measures, each centrality measure highlighting a different importance property of the network.

a) **Betweenness centrality**: Betweenness centrality quantifies the amount of information that flows through each parameter [14]. It is defined as the number of shortest paths that pass through a parameter, given by the following equation:

$$C_B(i) = \sum_{s \neq i \neq t} \frac{N_{st}(i)}{N_{st}}$$

Here, $N_{st}$ is the total number of shortest paths between the parameters, s and *t* and $N_{st}(i)$ is the total number of shortest paths between *s* and *t* that pass through *i*.

b) **Harmonic centrality:** Harmonic centrality of a parameter is defined as the sum of inverse of all geodesic distances (path lengths) from that parameter to all the other parameters[15]. Mathematically, harmonic centrality of an parameter, *i* is given by the following equation:

$$C_H(i) = \frac{1}{N-1} \sum_{\substack{s=1 \\ s \neq i}}^{N} \frac{1}{G(s,i)}$$

Here $G(s,i)$ is the geodesic distance or path length between the parameters *s* and *i*.

c) **Degree centrality:** Degree centrality or degree is defined as the total number of parameters, where each parameter is connected to in the complex network.

d) **Integrated centrality:** Each centrality measure signifies the importance of each parameter based on a pre-define characteristic. Where, the significance of each parameter across all centrality indices can be defined using integrated centrality [34] as shown in the following equation:

$$C_{int}(i) = \frac{1}{3}\left(\frac{C_B(i)}{\max(C_B)} + \frac{C_H(i)}{\max(C_H)} + \frac{C_D(i)}{\max(C_D)}\right)$$

Here, $C_{int}$ corresponds to the integrated centrality of the parameter, *i*; $C_B$, $C_H$ and $C_D$ represent the betweenness, harmonic and degree centralities of the parameter, *i* respectively and max () extracts the maximum values of each centrality measure across all the parameters.

Integrated Radiological Informatics System with comparison to Oncotype DX gene array23

*Integrated Radiological Informatics System with comparison to Oncotype DX gene array*Table 1 –Patient clinical and demographics

| Clinical | Mean | Standard Deviation |
|---|---|---|
| **Age** | 53 | 10 |
| **Tumor Grade (Elson)** | | |
| T1 | 8 | |
| T2 | 61 | |
| T3 | 12 | |
| **Phenotype** | | |
| ER+ | 81 | |
| PR+ | 74 | |
| Her2-Nu+ | 1 | |
| Triple Negative | 0 | |
| Ki67 (%) | 29 | 16 |
| **ODX** | | |
| Low | 19 | |
| intermediate | 50 | |
| high | 12 | |

**ER=Estrogen Receptor, PR-Progesterone Receptor, ODX=OncotypeDX,**

Table 2. Summary of the IRIS importance values of different informatic parameters.

| Informatics parameter | IRIS heatmap ranking |
|---|---|
| **Ki-67** | 1 |
| **ADC Ratio** | 2 |
| **K$^{trans}$** | 3 |
| **ADC Lesion** | 4 |
| **Percent initial uptake** | 5 |
| **Extra Vascular Fraction** | 6 |
| **Volume** | 7 |
| **Type of mass** | 8 |
| **ADC Glandular** | 9 |
| **Breast Density** | 10 |
| **Lesion Size** | 10 |
| **Margin** | 10 |
| **Lesion Type+Margin** | 10 |
| **Internal enhancement** | 10 |
| **Max Signal Intensity** | 10 |
| **Curve type** | 10 |
| **Time to Peak enhancement** | 10 |

Integrated Radiological Informatics System with comparison to Oncotype DX gene array

Table 3 Summary of the centrality values of different informatics parameters.

| Informatics parameter | Betweenness centrality | Harmonic centrality | Degree centrality | Integrated centrality |
|---|---|---|---|---|
| ADC Ratio | 46.00 | 1.49 | 5.00 | 12.82 |
| Curve type | 33.00 | 1.06 | 7.00 | 9.20 |
| Internal Enhancement | 29.00 | 0.75 | 3.00 | 8.08 |
| ADC Lesion | 17.00 | 1.47 | 5.00 | 4.74 |
| K$^{trans}$ | 16.00 | 0.95 | 6.00 | 4.46 |
| ADC Glandular | 15.00 | 0.73 | 3.00 | 4.18 |
| Ki-67 | 10.00 | 0.80 | 3.00 | 2.79 |
| Type of mass | 9.00 | 0.78 | 3.00 | 2.51 |
| Lesion Size | 7.00 | 0.65 | 3.00 | 1.95 |
| Breast Density | 5.00 | 0.66 | 3.00 | 1.39 |
| Lesion Type+Margin | 3.00 | 0.76 | 4.00 | 0.84 |
| Time to peak enhancement | 1.00 | 0.77 | 4.00 | 0.28 |
| Margin | 0.00 | 0.77 | 3.00 | 0.00 |
| Percent initial uptake | 0.00 | 0.85 | 3.00 | 0.00 |
| Max Signal Intensity | 0.00 | 0.63 | 3.00 | 0.00 |
| Volume | 0.00 | 0.59 | 3.00 | 0.00 |
| Extra Vascular Fraction | 0.00 | 0.82 | 3.00 | 0.00 |

**ADC=Apparent Diffusion Coefficient**

**Figures**

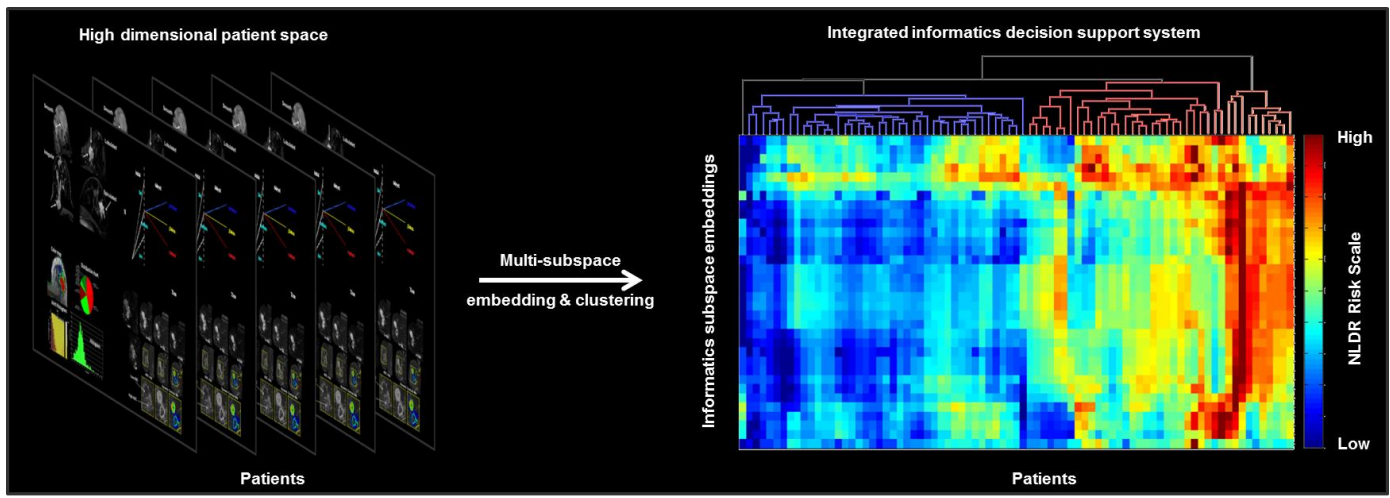

Figure 1. Illustration of the multi-subspace embedding and clustering method. The high dimensional patient space (left) consisting of different patients and their corresponding clinical and imaging parameter information is transformed into an integrated radiomics informatics system (IRIS) decision support system (right) using the multi-subspace embedding and clustering method. The IRIS results are represented using a heatmap where color scale (blue – red) indicates risk identified by each embedding while the hierarchical tree structure indicates the final patient classes (low, intermediate and high-risk)

*Integrated Radiological Informatics System with comparison to Oncotype DX gene array*

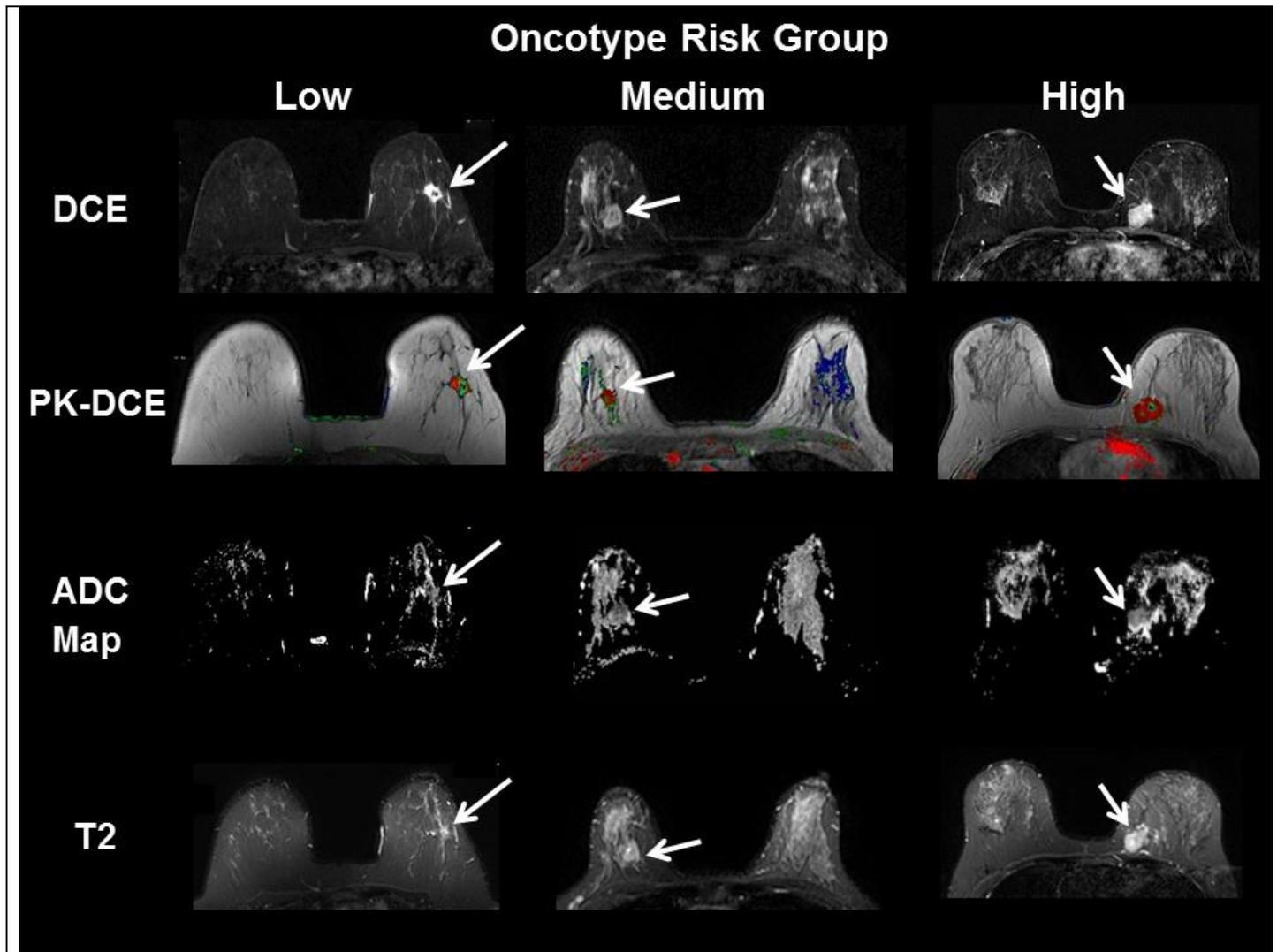

Figure 2. Demonstration of multiparametric breast MRI imaging of each risk group defined by the Oncotype DX. **Left Column**) typical imaging of the low risk patient. **Middle Column**) typical imaging of the medium risk patient, and **Right Column**) typical imaging of a high-risk group patient. Note, the PK-DCE all demonstrate malignant phenotype, however, by integrating all the data using IRIS, we were able to separate each Oncotype DX group.

28

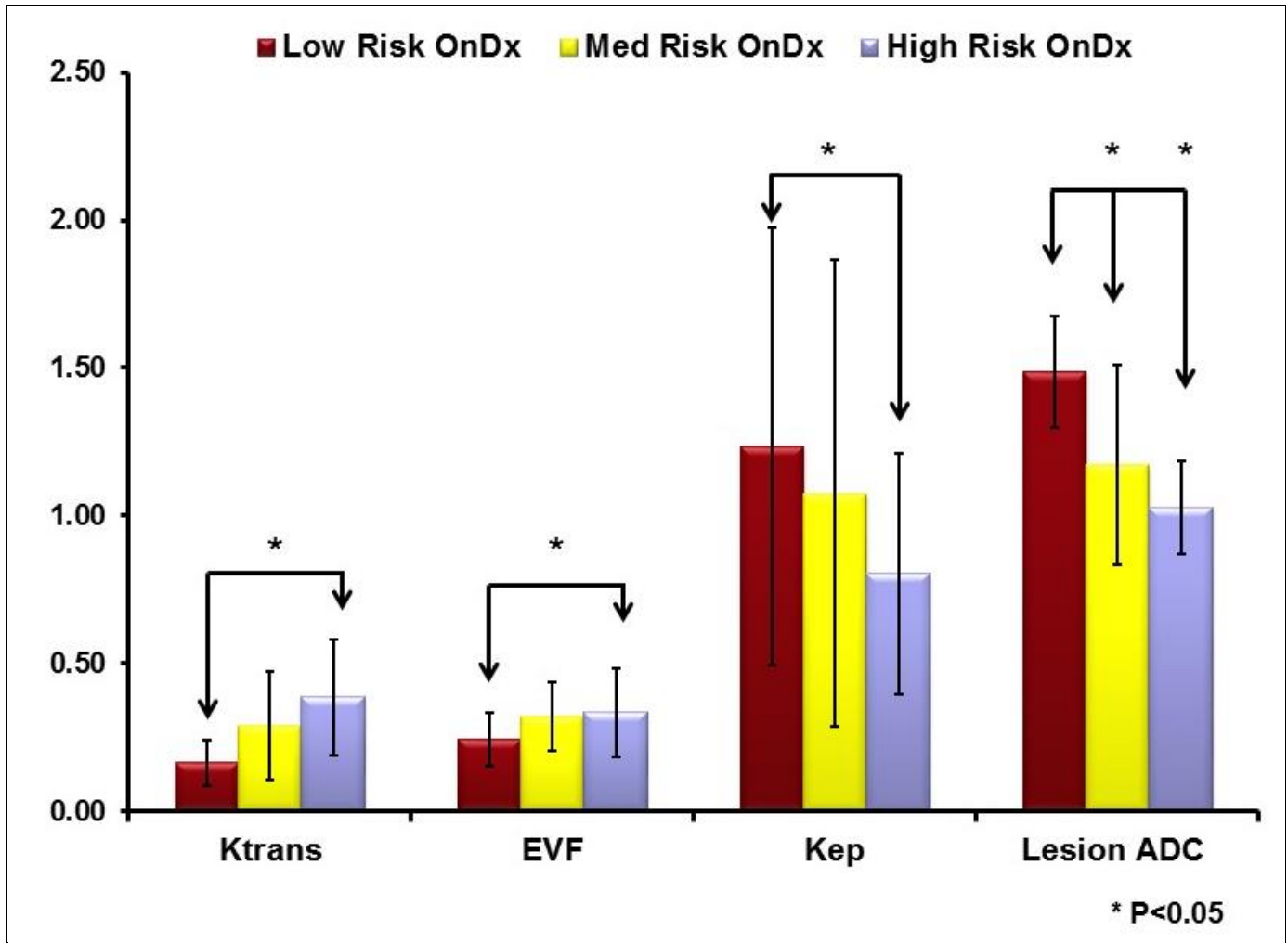

Figure 3. Bar graphs of quantitative multiparametric MRI parameters from the NLDR model. There are significant differences between each group of patients.

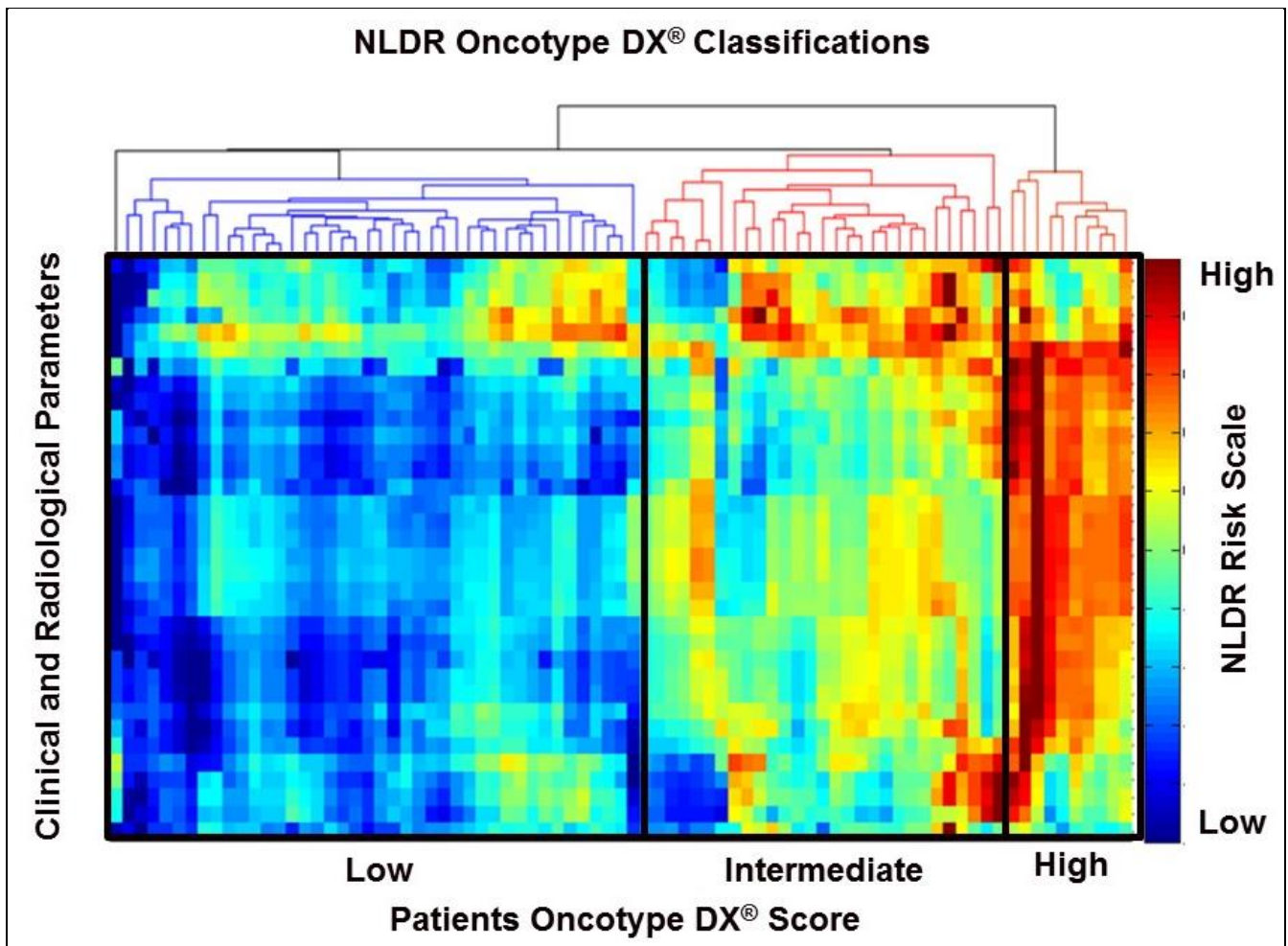

Figure 4. The heat map demonstrates the stratification of the radiological parameters and risk groups correlated with the OncotypeDX score. Note, there are high risk factors perceived in the low risk that may not been noted in clinical evaluations.

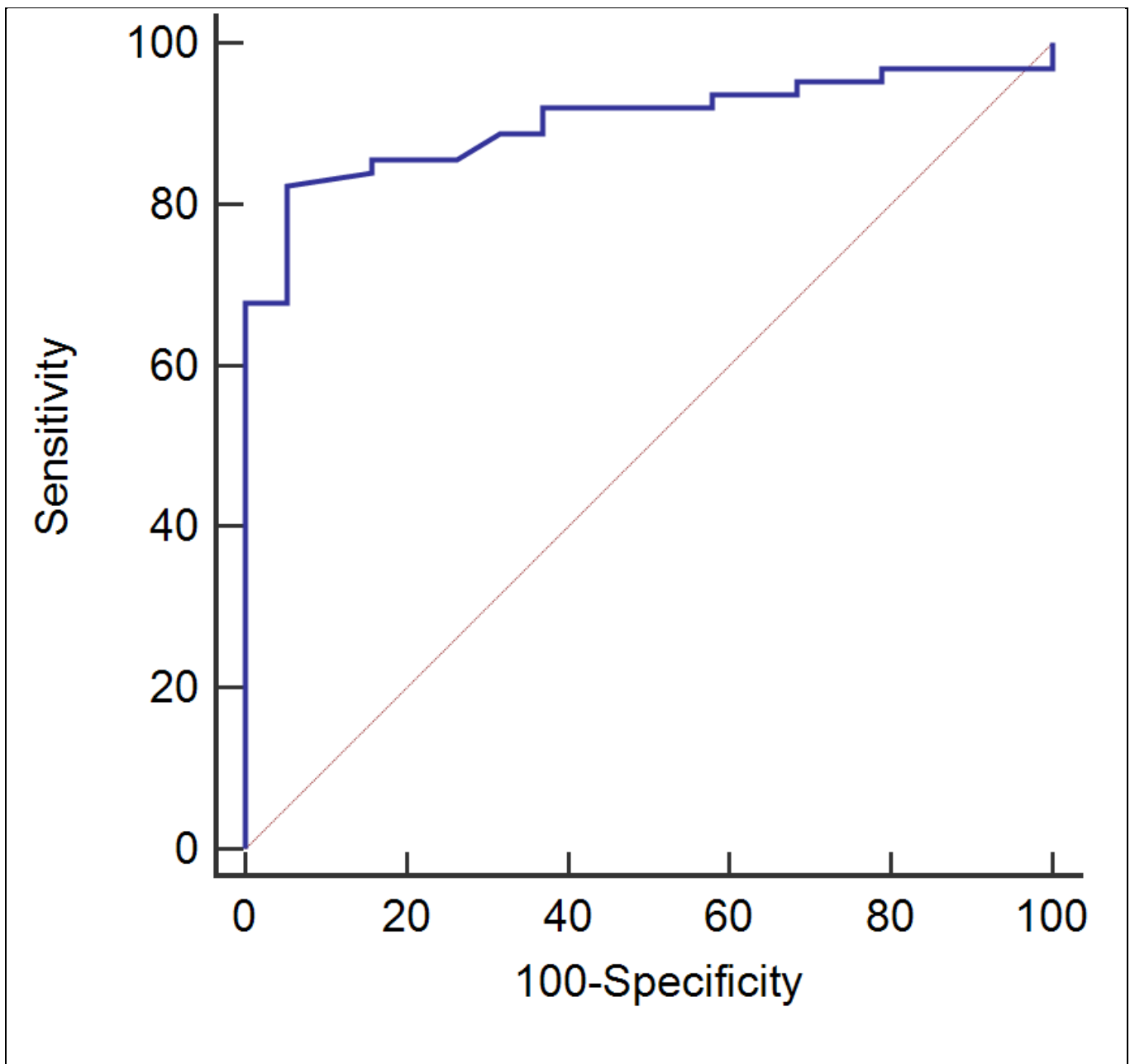

Figure 5. The Area under the Curve (AUC) graph of the top risk predictors in the NLDR model with the OncotypeDX score. The sensitivity was 95% and specificity was 82% with an AUC =0.90 (95% CI = 0.83-0.97).

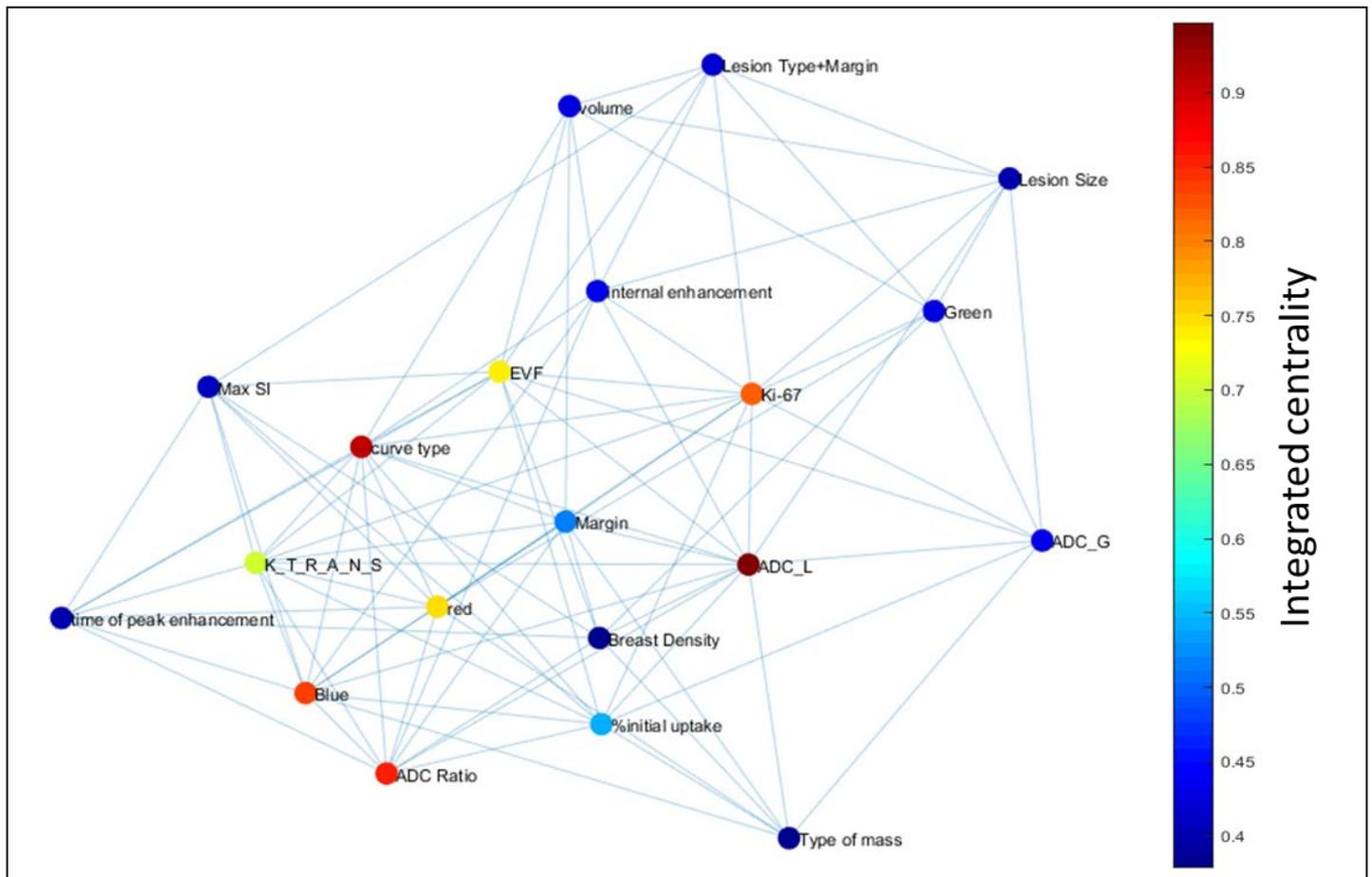

Figure 6. Visualization of integrated centrality on the complex network. The nodes were color coded from blue to red with blue representing minimum and red representing maximum integrated centrality values. The top IRIS parameters demonstrate high integrated centrality metrics demonstrating their importance.